\documentclass[10pt, conference, compsocconf]{IEEEtran}
\pdfoutput=1
\IEEEoverridecommandlockouts
\usepackage{cite}
\usepackage{amsmath,amssymb,amsfonts}
\usepackage{algorithmic}
\usepackage{graphicx}
\usepackage{textcomp}
\usepackage{xcolor}
\usepackage{booktabs}
\usepackage{subcaption}
\usepackage{multirow}
\def\BibTeX{{\rm B\kern-.05em{\sc i\kern-.025em b}\kern-.08em
T\kern-.1667em\lower.7ex\hbox{E}\kern-.125emX}}
\begin{document}

\title{Social Isolation and Serious Mental Illness: The Role of Context-Aware Mobile Interventions}

\makeatletter
\newcommand{\linebreakand}{%
  \end{@IEEEauthorhalign}
  \hfill\mbox{}\par
  \mbox{}\hfill\begin{@IEEEauthorhalign}
}
\makeatother

\author{
  \IEEEauthorblockN{Subigya Nepal\IEEEauthorrefmark{1}}
  \IEEEauthorblockA{
    \textit{Dartmouth College}\\
    Hanover, NH, USA \\
    sknepal@cs.dartmouth.edu}
  \and
  \IEEEauthorblockN{Arvind Pillai\IEEEauthorrefmark{1}}
  \IEEEauthorblockA{
    \textit{Dartmouth College}\\
    Hanover, NH, USA \\
    arvind.pillai.gr@dartmouth.edu}
  \and
  \IEEEauthorblockN{Emma M. Parrish}
  \IEEEauthorblockA{\textit{San Diego State University} \\
    \textit{University of California San Diego}\\
    San Diego, CA, USA \\
emparris@health.ucsd.edu}
  \linebreakand 
  \linebreakand 
  \linebreakand 
  \IEEEauthorblockN{Jason Holden}
  \IEEEauthorblockA{
    \textit{University of California San Diego}\\
    San Diego, CA, USA \\
   jlholden@health.ucsd.edu}
  \and
  \IEEEauthorblockN{Colin Depp}
  \IEEEauthorblockA{
    \textit{University of California San Diego}\\
    San Diego, CA, USA \\
    cdepp@health.ucsd.edu}
     \and
  \IEEEauthorblockN{Andrew T. Campbell}
  \IEEEauthorblockA{
    \textit{Dartmouth College}\\
    Hanover, NH, USA \\
    andrew.t.p.campbell@gmail.com}
 \linebreakand
      \IEEEauthorblockN{Eric Granholm}
  \IEEEauthorblockA{
    \textit{University of California San Diego}\\
    San Diego, CA, USA \\
    egranholm@health.ucsd.edu}
}


\maketitle
\def\thefootnote{*}\footnotetext{These authors contributed equally to this work.}
\begin{abstract}
Social isolation is a common problem faced by individuals with serious mental illness (SMI), and current intervention approaches have limited effectiveness. This paper presents a blended intervention approach, called mobile Social Interaction Therapy by Exposure (mSITE), to address social isolation in individuals with serious mental illness. The approach combines brief in-person cognitive-behavioral therapy (CBT) with context-triggered mobile CBT interventions that are personalized using mobile sensing data. Our approach targets social behavior and is the first context-aware intervention for improving social outcomes in serious mental illness.
\end{abstract}


\section{Introduction}
Mental health is a pressing public health concern affecting a significant portion of the population in the United States. Approximately one in five adults in the country experience mental illness each year. Among individuals with serious mental illness (SMI) (e.g., schizophrenia, bipolar disorder), social isolation is a pervasive issue that negatively impacts quality of life, functional ability and mental health outcomes. For example, recent studies have shown that over 75\% of people with psychotic disorders report loneliness. Moreover, studies have shown that social isolation can lead to physical health problems, such as, an increased risk of cardiovascular disease and mortality. In fact, social isolation is a severe problem with healths  comparable to smoking and stroke combined. Unfortunately, the worldwide rate of social isolation is increasing, and those with SMI are at particularly high risk. Studies have found that patients who are more socially isolated have more severe negative symptoms, depression, and worse social functioning. Addressing social isolation is critical for treatment response and recovery, yet there is a shortage of interventions targeting social isolation in SMI.

Mobile technology has emerged as a potential tool for addressing mental health problems. Mobile sensing technology enables the collection of data on various aspects of an individual's life, such as physical activity, sleep patterns, and social interactions. This data can then be used to develop personalized interventions that target specific mental health problems, such as social isolation. Mobile-based interventions, including cognitive-behavioral therapy (CBT) programs, mindfulness apps, have shown promise in addressing a range of mental health issues, including depression and anxiety. While interventions for mental health have largely focused on addressing symptoms, there remains a significant gap in addressing social functioning and its impact on overall well-being. 


Research suggests that in psychotic disorders, social isolation is determined by both reduced approach and avoidance mechanisms. The approach mechanism encourages people to engage in actions that are rewarding or pleasurable, while the avoidance mechanism motivates people to avoid situations or actions that are threatening or negative. Individuals with schizophrenia, for example, may experience decreased motivation to socialize and have diminished reward learning from social interactions, which is worsened by negative symptoms. At the same time, anxiety and social threat perception may result in social avoidance. Interventions aimed at increasing motivation for socialization or reducing social anxiety have had modest success in treating negative symptoms of schizophrenia. However, most of such interventions are intensive and are not scalable. Previous studies have shown that both diminished reward and anxious avoidance play a role in determining social interactions in individuals with schizophrenia. These studies have also found that context plays an important role in social behavior, with different thoughts and emotions being present when interacting with others (negative defeatist appraisals) versus being alone at home (anxious threat about leaving)~\cite{Depp2019, Parrish2020, Depp2016, Granholm2013}. The premise of our study, therefore, is that, a context-aware CBT intervention that targets specific thoughts and emotions related to social behavior in different contexts may be more effective in improving social motivation and functioning in individuals with serious mental illness.

\begin{figure*}
    \centering
    \includegraphics[width=\linewidth]{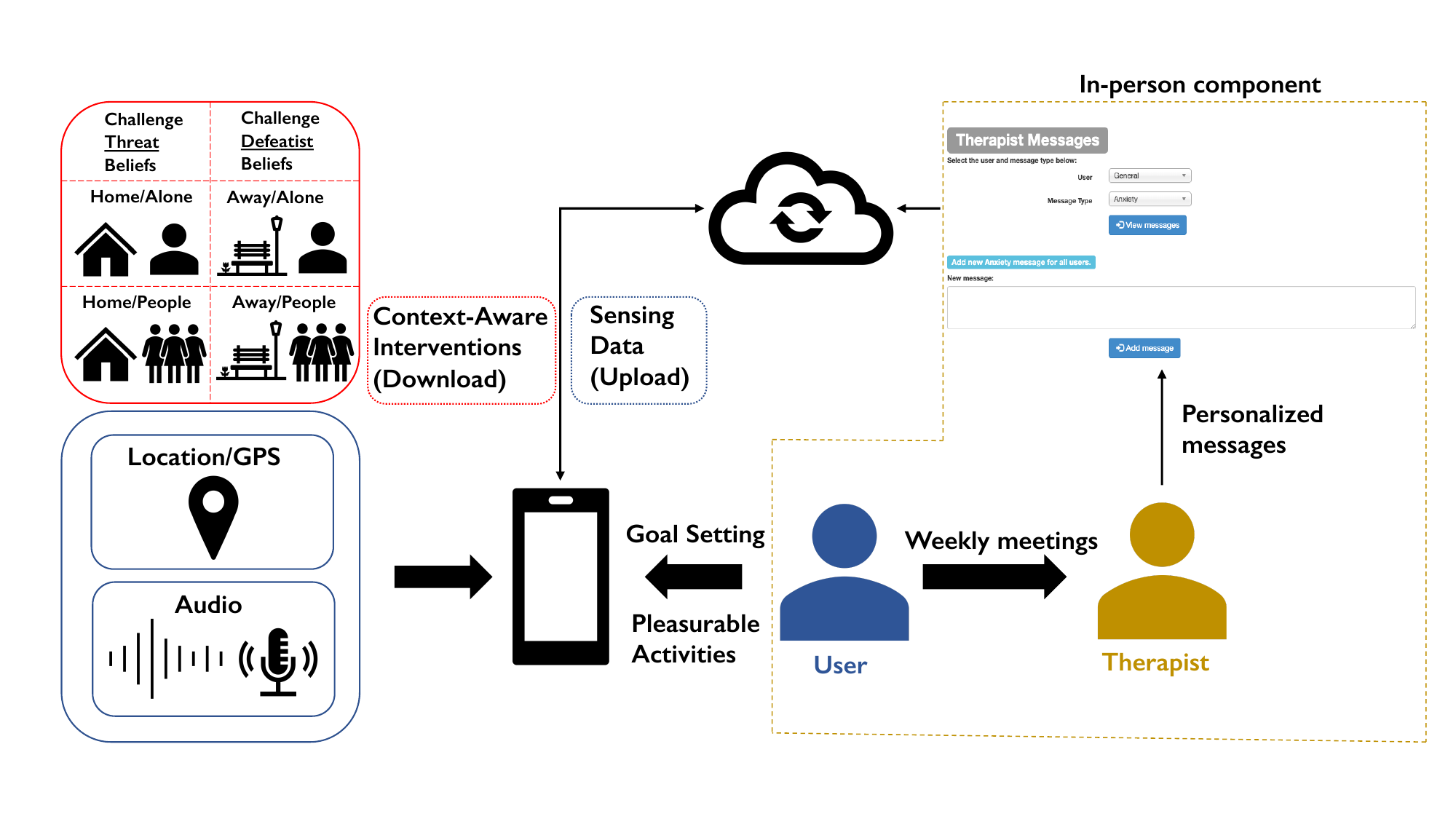}
    \caption{mSITE's blended CBT approach using context-triggered mobile interventions and therapist-in-the-loop personalization.}
    \label{fig:msiteworkflow}
\end{figure*}

We propose a novel blended intervention approach that combines brief in-person CBT with context-triggered mobile CBT interventions to address social isolation in SMI. The intervention is personalized, participatory, and employs a novel and flexible architecture that facilitates intervention based on sensor-detected context. It is also the first context-aware intervention for improving social outcomes in SMI. The blended intervention approach is designed to target social behavior, rather than just symptoms. The intervention employs brief in-person CBT to set socialization goals, introduce the cognitive model and thought challenging skills to address defeatist attitudes and social threat beliefs, and plan pleasurable social activities for the mobile component of the intervention. The mobile component consists of context-triggered CBT interventions that aim to increase social behavior through reminders, reinforcement, activity pleasure savoring, and cognitive restructuring. The mobile component is personalized based on the data collected through mobile sensing technology  (specifically data on the patient's location and social interactions) as well as with the help of personalized messages entered by the therapists. Furthermore, the architecture of our intervention is flexible, enabling dynamic modification of intervention strategies based on the patient's context, paving the way for future “Just-in-time” and adaptive interventions. Note that this manuscript represents initial results and insights from an ongoing study. Full trial results, as well as a greater focus on the therapy aspect of the intervention, will be reported in a later paper.The aims of this manuscript  are as follows:

\begin{enumerate}
    \item We describe the details of our study and the design of mobile-based tools that captures novel sensing modalities (e.g., conversation-mobility sensor using gps and microphone).
    \item We present a representative case study to illustrate the effectiveness of context-aware mobile interventions for improving social functioning
    \item We present the implications of our work and elucidate extensions that broadly benefit the mobile sensing and mental health community.
\end{enumerate}

\section{RELATED WORK}
Mobile-based mental health interventions have exhibited potential for providing real-time support to individuals. For instance, \cite{granholm2012mobile} observed that longitudinal text message-based interventions improve social isolation in individuals with schizophrenia or schizoaffective disorder. Social isolation is common among older adults \cite{choi2012computer, balki2022effectiveness}. Consequently, \cite{choi2012computer} investigated the effectiveness of technology-based interventions for loneliness in an aging population. Their results suggest that tools like social networks and instant text messages lower the incidence of isolation and feelings of loneliness. Although various studies have employed CBT algorithms in mobile apps to address mental health symptoms and enhance social interactions and self-efficacy beliefs \cite{Depp2018, BenZeev2014, BenZeev2013, Schlosser2016, Bucci2018}, no existing context-aware systems have specifically focused on social behavior in severe mental illness (SMI). Blended interventions, which merge mobile and in-person therapies while reducing in-person contact by 50-90\%, have been more effective than app-only approaches \cite{AlvarezJimenez2014, BenZeev2014, BenZeev2013}. One recent pilot trial of a blended intervention for psychosis used coping-focused sessions for voices blended with an app and found improved coping and trend-level reduction in voices~\cite{Bell2020}. However, it focused on symptoms, and there have been limited blended interventions for social isolation tailored to SMI.  We identified only one brief 3-week intervention providing feedback of ecological momentary assessment (EMA) responses and behavior change suggestions, which showed improvement in symptoms but not social functioning~\cite{Hanssen2020}. Our study presents a novel contribution to the field by introducing a personalized, blended context-aware mobile intervention aimed at enhancing social functioning in individuals with SMI. This innovative approach specifically targets social behavior, expanding on previous work and addressing a gap in context-aware systems for SMI populations.

\section{METHODOLOGY}
\subsection{Participant Selection}
Our study aims to recruit a diverse sample of 50 participants aged 18-65 with SMI, including schizophrenia, schizoaffective, or bipolar I disorder with psychosis. In this manuscript, we report on the preliminary data on feasibility and app performance in N=5 participants with schizophrenia. 75\% of the participants we have enrolled are white (N=3), 50\% are male (N=2) and their mean age and mean years of education are 54.8 years (Range=47-59) and 12 respectively. Inclusion criteria of our study include a minimum level of social avoidance, defined as a score of $\geq$ 2 on the Scale for Assessment of Negative Symptoms asociality item, a reading level of at least 6th grade on the Wide Range Achievement Test-4 Reading subtest, and stability in medications (no hospitalizations or medication changes in the 4 months prior to enrollment).
Exclusion criteria include prior receipt of CBT in the past 2 years, greater than moderate disorganization on the Positive and Negative Symptom Scale, and alcohol or substance dependence in the past 3 months based on DSM-5 criteria. In addition, participants requiring a higher level of care (e.g., hospitalization, severe medical illness) are excluded, as well as those who are unable to adequately see or manually manipulate a mobile phone, since ours is a blended intervention that uses a mobile device. 

\subsection{Study Design}

\begin{figure*}[t]
     \centering
     \begin{subfigure}[b]{0.24\textwidth}
         \centering
         \includegraphics[width=0.8\textwidth]{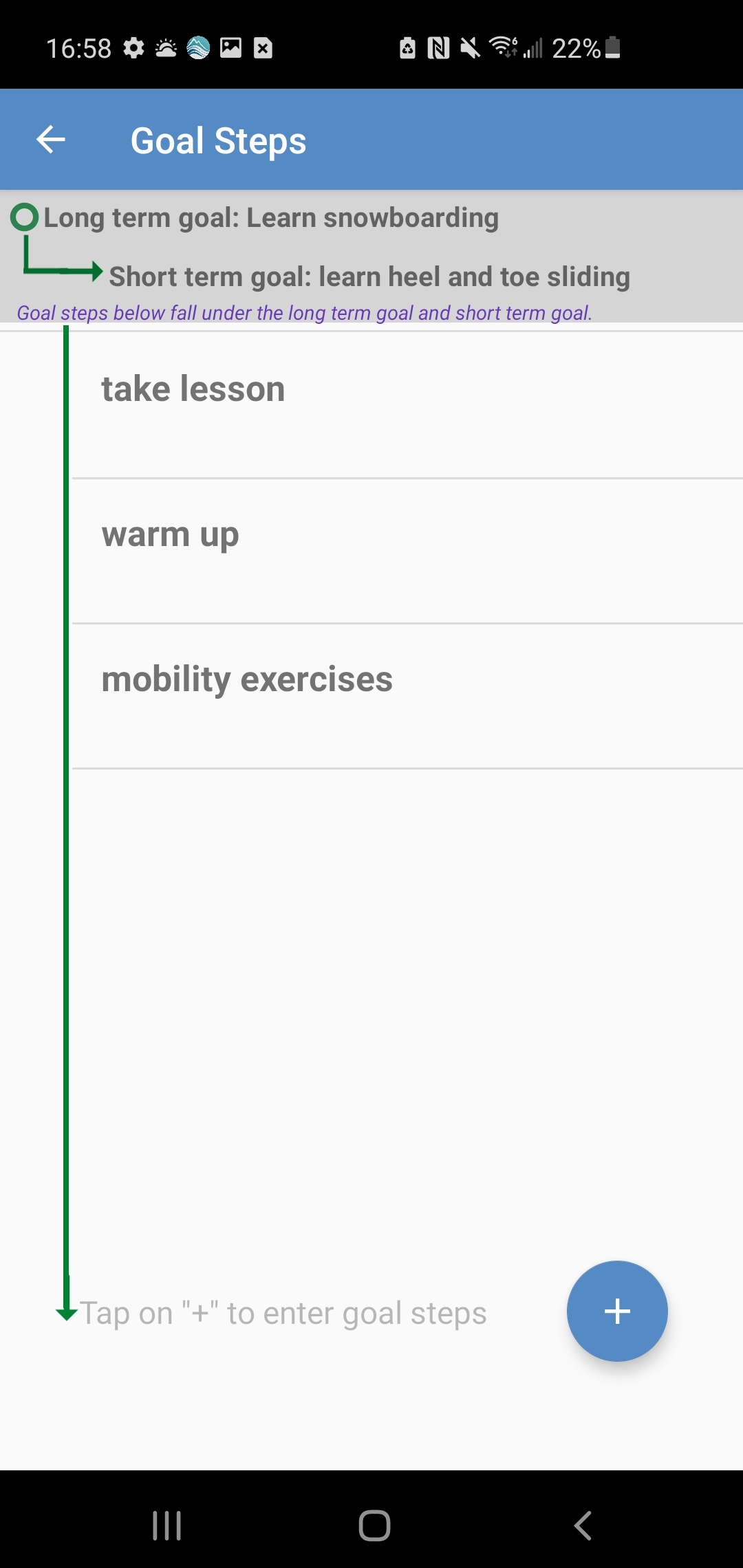}
         \caption{}
         \label{fig:goalsteps}
     \end{subfigure}
     \hfill
     \begin{subfigure}[b]{0.24\textwidth}
         \centering
         \includegraphics[width=0.8\textwidth]{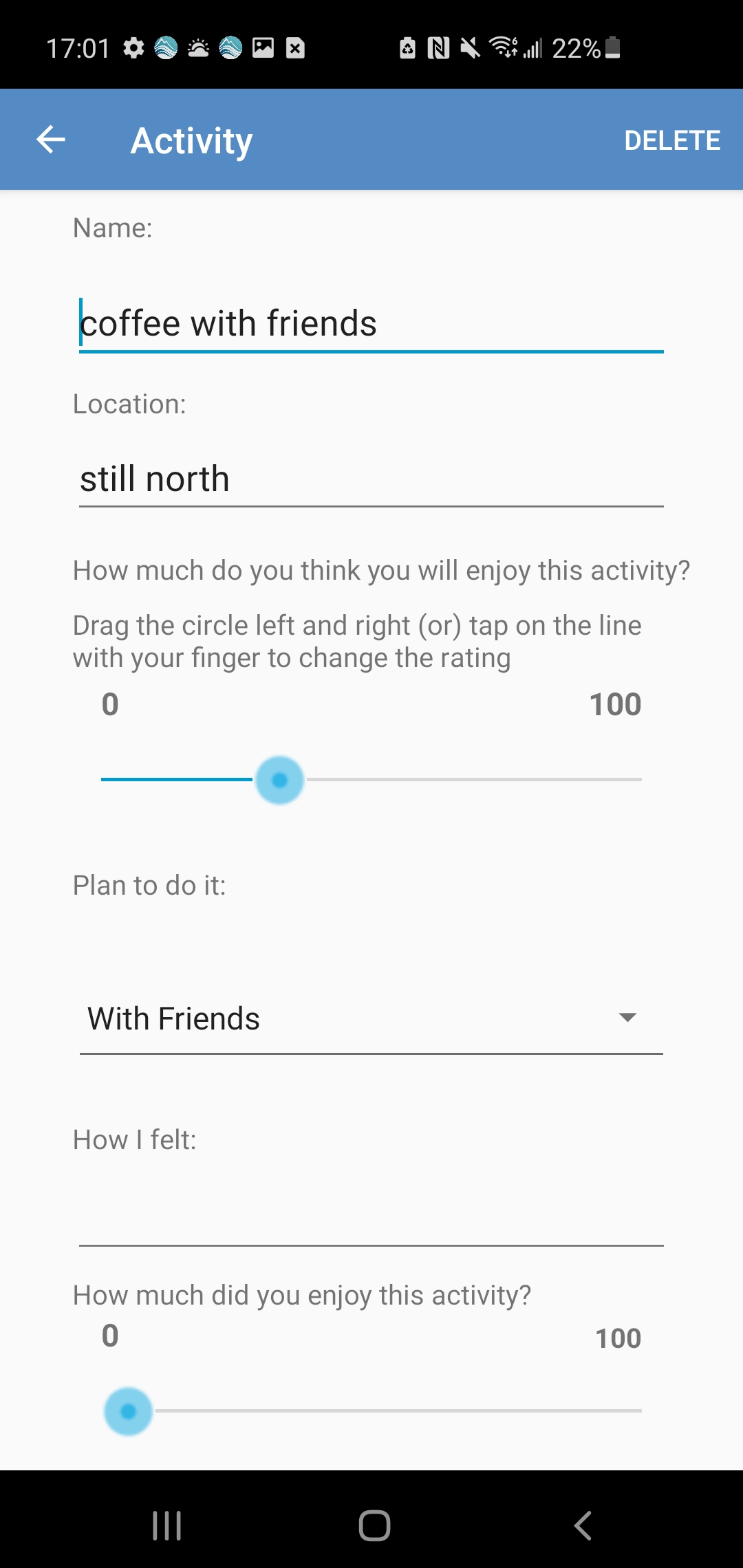}
         \caption{}
         \label{fig:activitypre}
     \end{subfigure}
     \hfill
     \begin{subfigure}[b]{0.24\textwidth}
         \centering
         \includegraphics[width=0.8\textwidth]{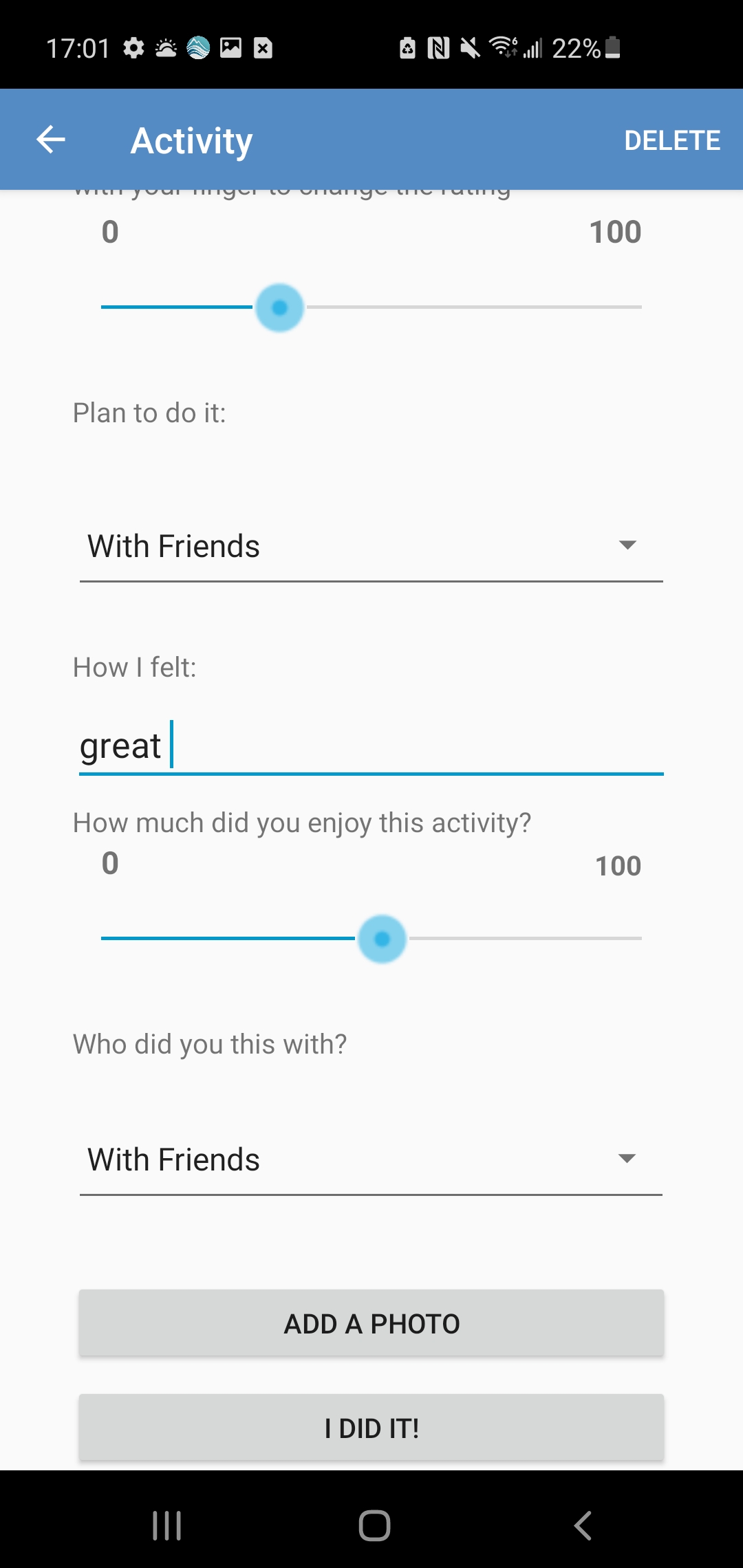}
         \caption{}
         \label{fig:activitypost}
     \end{subfigure}
    \begin{subfigure}[b]{0.24\textwidth}
         \centering
         \includegraphics[width=0.8\textwidth]{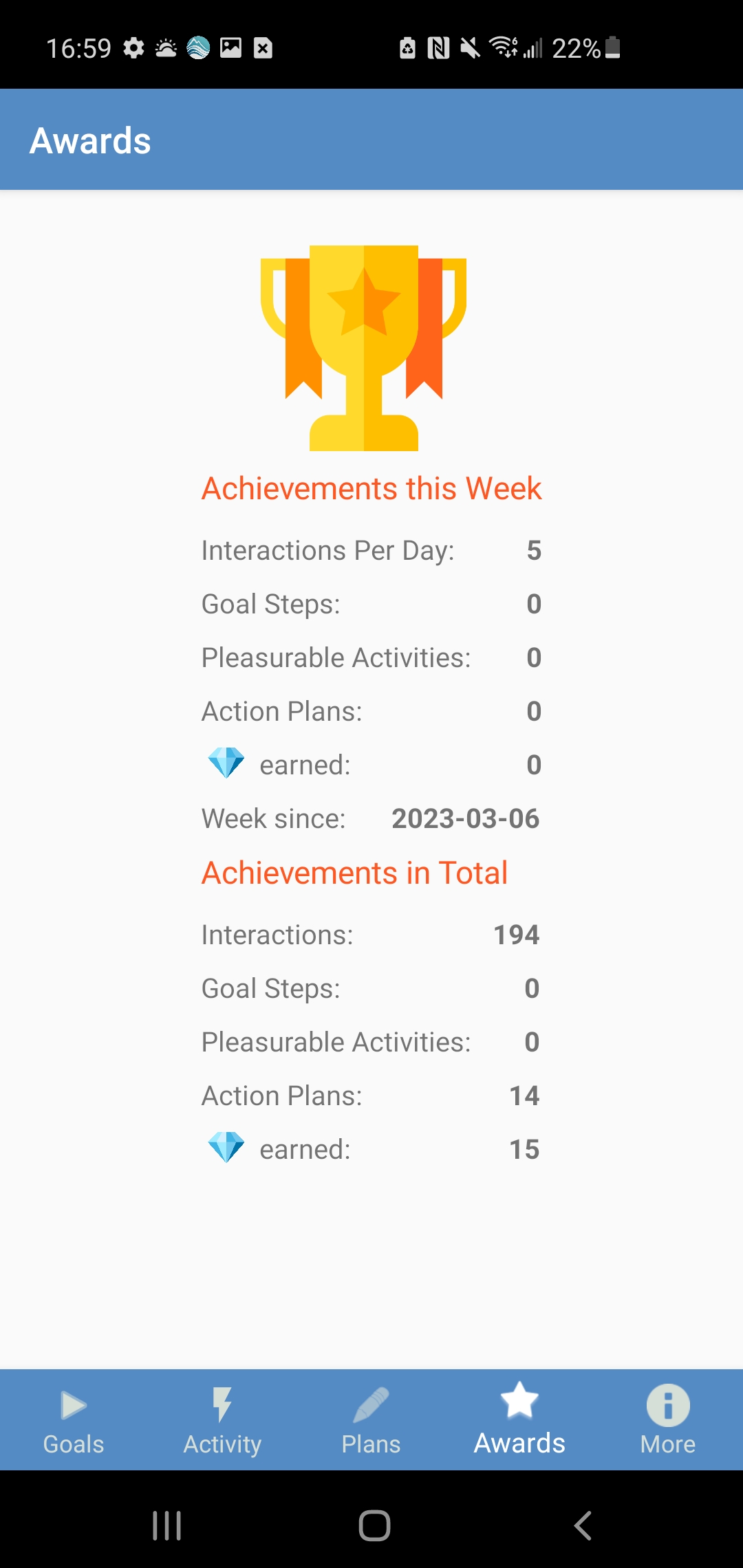}
         \caption{}
         \label{fig:awards}
     \end{subfigure}
      \caption{Important mSITE screens: (a) User entered goal steps in a hierarchical view, i.e., Top-level long term goals, contains short-term goals, which contains goal steps, (b) User enters a pleasurable social activity for the day, (c) User describes how their pleasurable activity went using freeform text, sliders, and can add photos, (d) Gamification using awards the user has accumulated through the study.}
        \label{fig:msitescreens}
\end{figure*}

The study involves a brief blended intervention that combines in-person and mobile CBT to reduce social isolation. Social isolation can be caused by a lack of reward from relationships and avoiding social situations due to anxiety. When someone is alone, they might start to have negative thoughts about social situations and then avoid them, which only reinforces their negative beliefs. Modifying negative beliefs and behaviors when alone can help someone feel more comfortable and engaged in social situations, which can lead to positive changes in their thoughts, emotions, and social behavior. Our hypothesis is that the use of our mobile-assisted CBT intervention, called mobile Social Interaction Therapy by Exposure (mSITE), will lead to an increase in social interactions among individuals with SMI. This increase in social interactions will in turn lead to a reduction in the severity of experiential negative symptoms and ultimately improve social functioning. Note that the study is approved by the Institutional Review Board of University of California, San Diego and Dartmouth College. It is also registered as a clinical trial in clincaltrials.gov.

We are conducting a 24-week open pilot trial of mSITE. Our trial focuses on feasibility, acceptability, and initial indication of improvement in our target, i.e., social activity. The intervention involves eight weekly 1-hour in-person sessions combined with mobile app use, followed by weekly 15-minute remote coaching calls with mobile app use for an additional 16 weeks, for a total of 24 treatment weeks. We have included weekly remote coaching, as the literature suggests that mobile interventions are more effective when coaching is included. Our assessments will be carried out at baseline, 8 weeks (end of in-person treatment), 12 weeks, 18 weeks, and 24 weeks, with one-week bursts of ecological momentary assessments (EMA) surveys at each of these weeks. Our prior research has demonstrated that improvements in outcomes can be achieved in 24 weeks or fewer. However, if we can attain our target engagement at the earliest possible assessment point, which could be at week 12 or 18, it would allow us to create a shorter yet efficient intervention. Effectively strengthening and shortening intensive psychosocial interventions could help overcome implementation barriers that arise due to the high cost and burden of treatment. In addition, it could improve access to evidence-based practices for individuals who are underserved by the current mental healthcare system.

\subsection{In-person Intervention Component}
The intervention we are using has eight sessions and is based on different evidence-based treatments for mental health issues, including CBT. During the sessions, masters-level therapists, comparable to therapists in a community mental health system, work with patients to develop skills to help them achieve their goals for recovery. We start by setting a recovery goal, then we teach patients how to use a thought challenging skill called the 3Cs (Catch-It, Check-It, Change-It; where \textit{It} is a thought) to address defeatist beliefs and social avoidance and work towards their goals. 3Cs is a CBT technique used to help individuals identify and challenge dysfunctional thoughts and replace them with more helpful, accurate ones. We teach patients how to challenge defeatist attitudes, social threats, and avoidance behaviors using CBT skills. We also educate patients about how their thoughts, actions, and feelings are related, and use experiments and exposure to help them change dysfunctional beliefs. The intervention includes tests of expectations in social interactions to help patients achieve their goals. Essentially, this means that the intervention uses real-life social interactions as a way to help patients challenge and modify their beliefs and behaviors.

\subsection{Mobile Intervention Component}

The mobile intervention component uses a smartphone application (referred to as the mSITE app) to passively collect sensing data, e.g., activity, GPS, audio, light, phone usage, application usage, etc. The GPS sensors and microphones are used to detect if the user is home or away and around conversation, respectively. Moreover, the app alerts the users three times daily with personalized CBT scripts using information from the sensors and therapists. Note that the mSITE app does not record any raw conversation (only whether conversation was present or not and the amplitude of sound). While we collect other passive sensing data, only the GPS and conversations related information are used to drive interventions. Other passive sensing data are used only for analysis purposes.

\begin{figure*}
     \centering
     \begin{subfigure}[b]{0.16\textwidth}
         \centering
         \includegraphics[width=0.8\textwidth, scale=0.5]{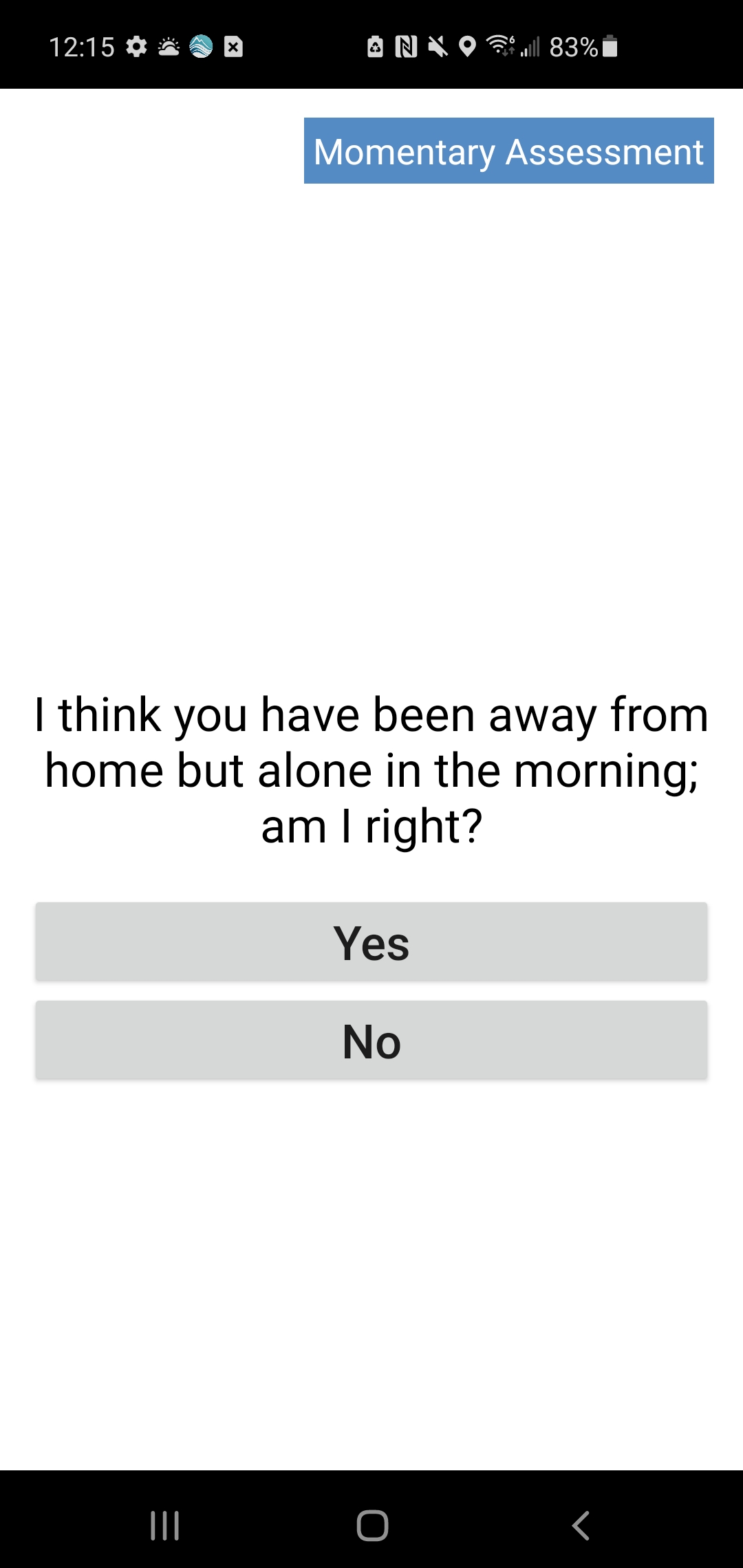}
         \caption{}
         \label{fig:awayalone1}
     \end{subfigure}
     \hfill
     \begin{subfigure}[b]{0.16\textwidth}
         \centering
         \includegraphics[width=0.8\textwidth, scale=0.5]{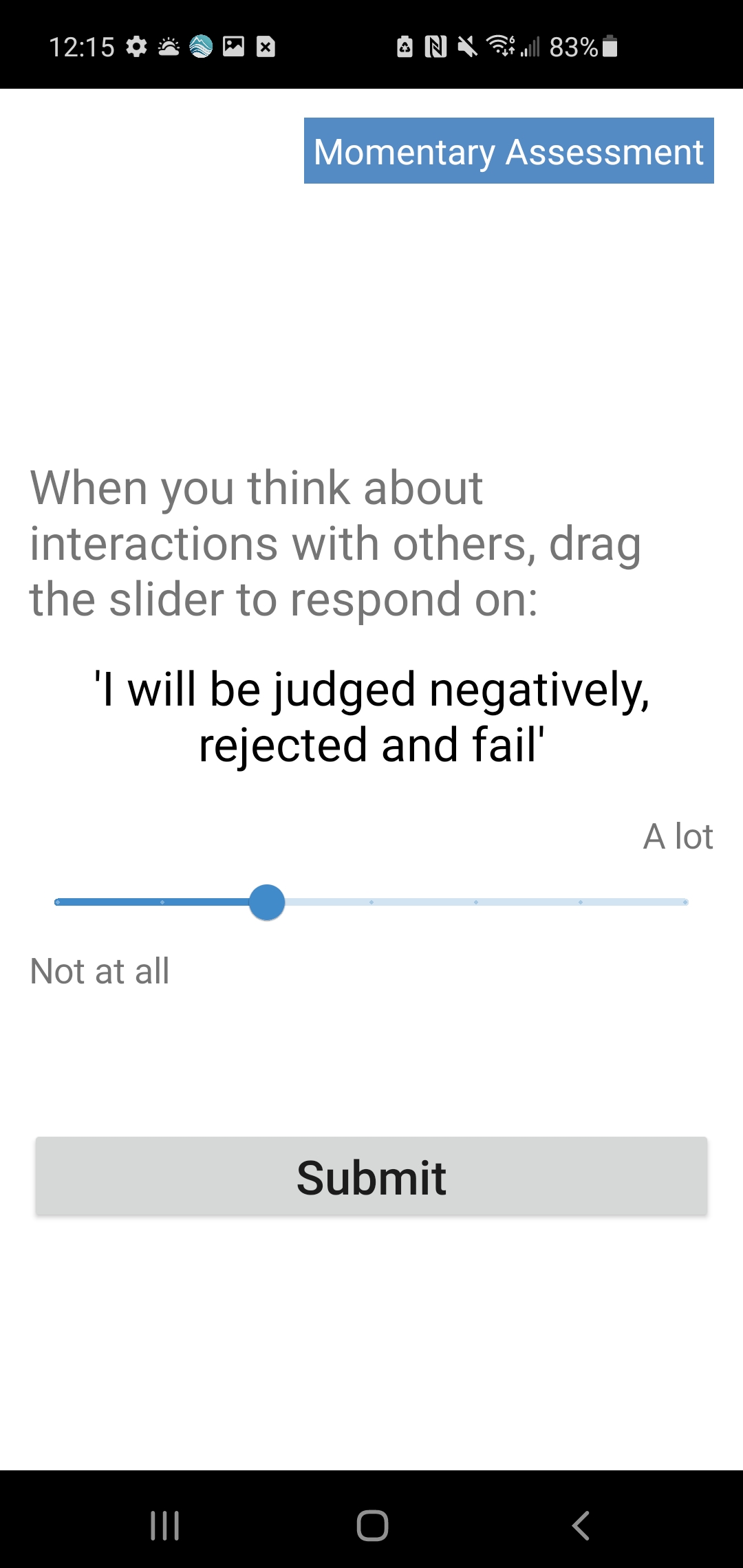}
         \caption{}
         \label{fig:awayalone2}
     \end{subfigure}
     \hfill
     \begin{subfigure}[b]{0.16\textwidth}
         \centering
         \includegraphics[width=0.8\textwidth, scale=0.5]{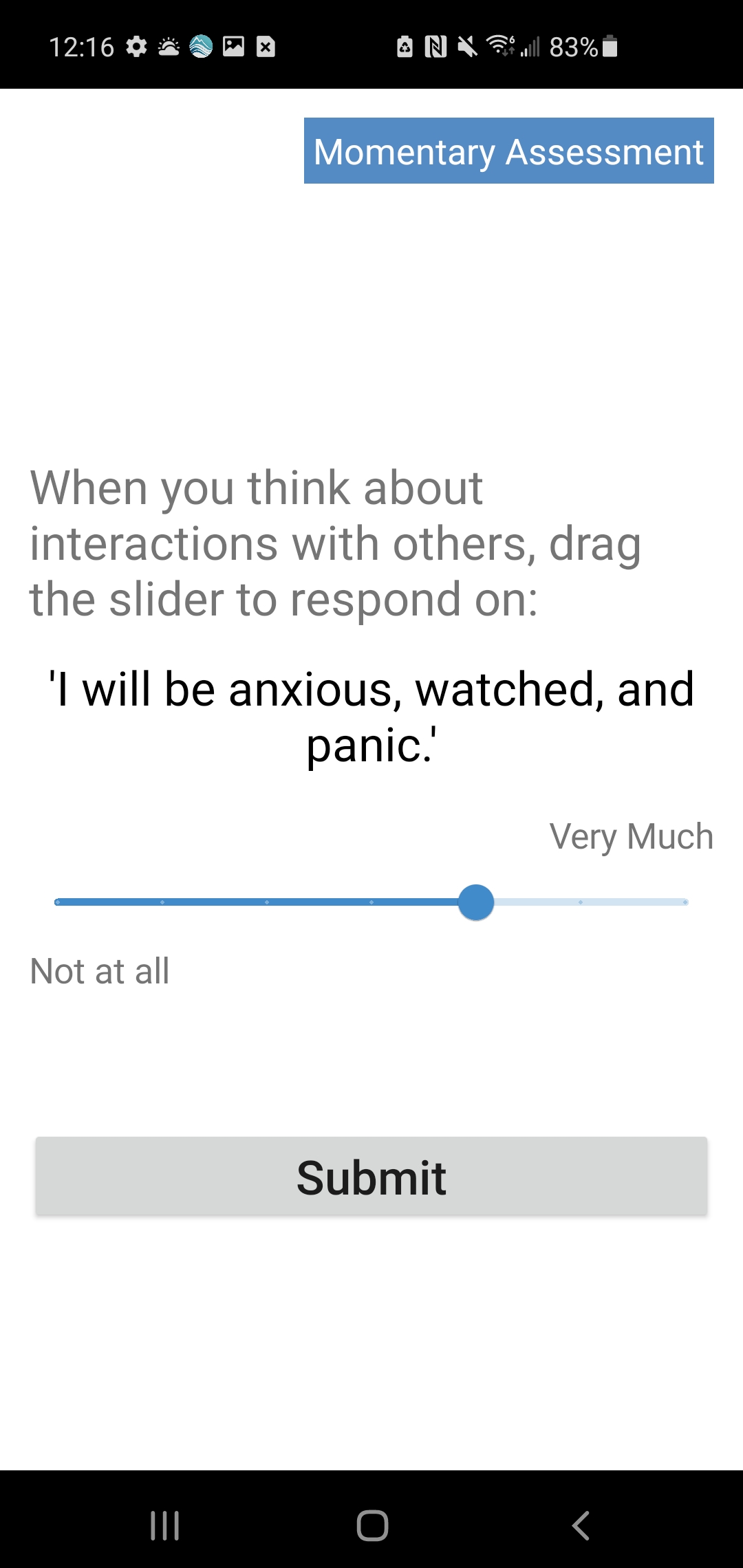}
         \caption{}
         \label{fig:awayalone3}
     \end{subfigure}
    \begin{subfigure}[b]{0.16\textwidth}
         \centering
         \includegraphics[width=0.8\textwidth]{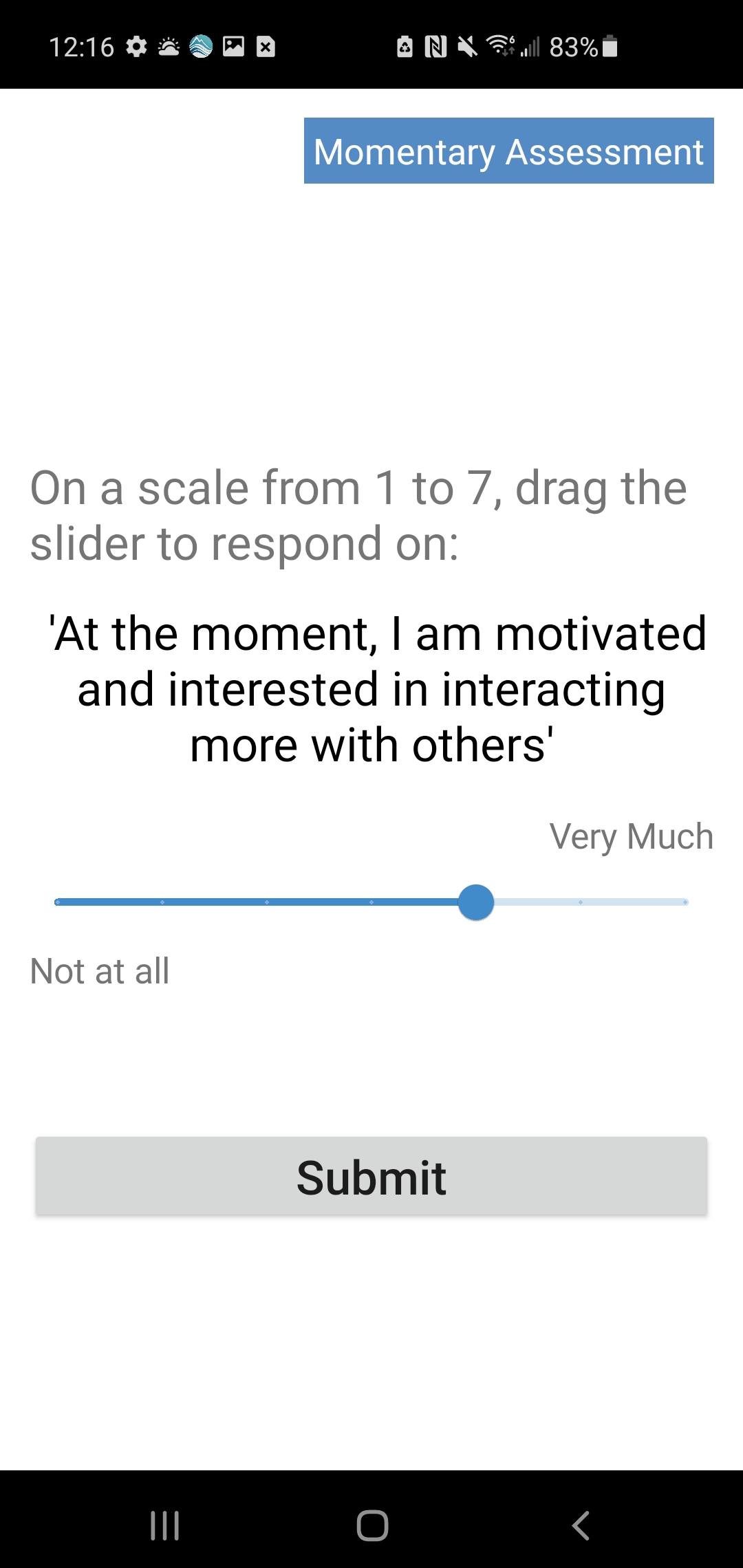}
         \caption{}
         \label{fig:awayalone4}
     \end{subfigure}
    \begin{subfigure}[b]{0.16\textwidth}
         \centering
         \includegraphics[width=0.8\textwidth, scale=0.5]{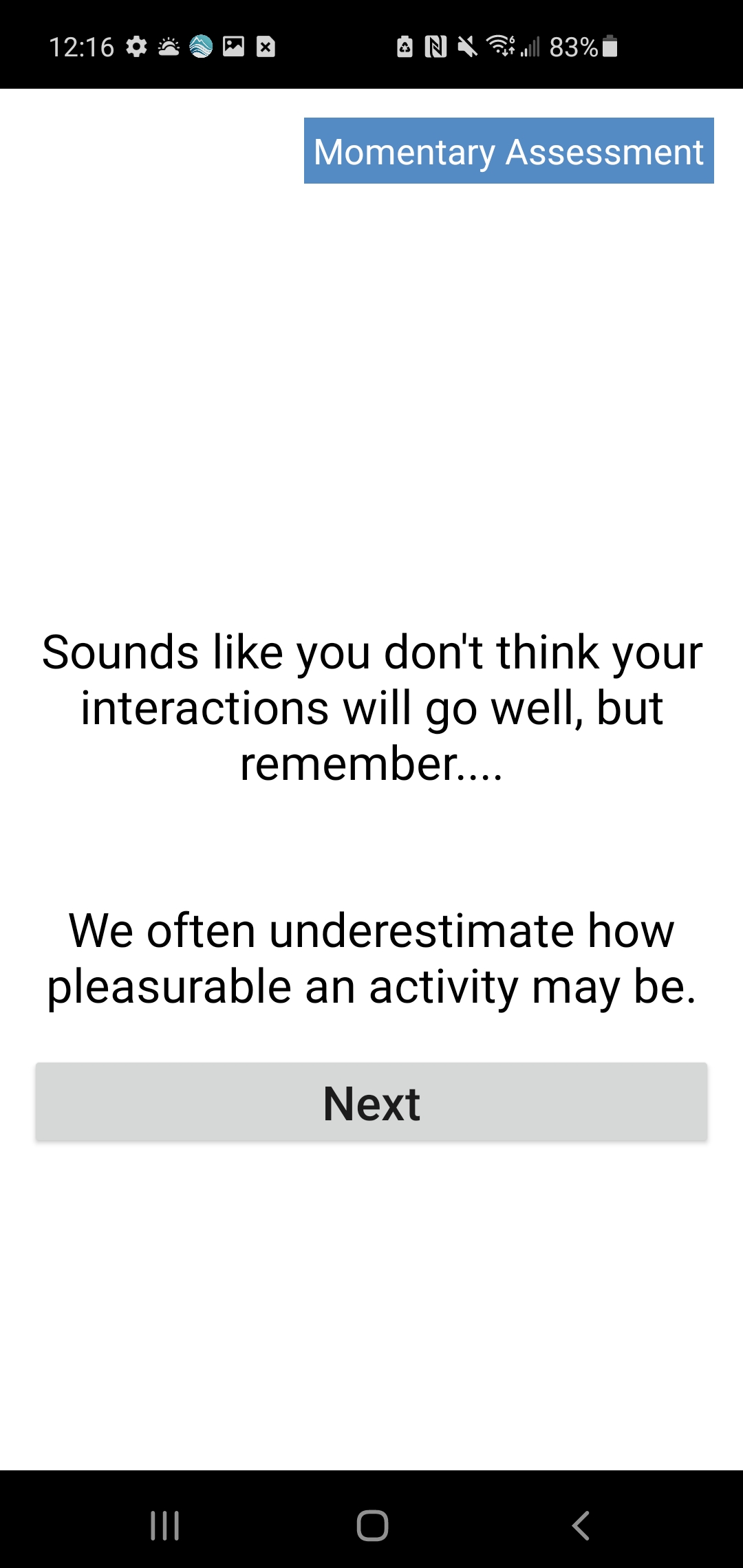}
         \caption{}
         \label{fig:awayalone5}
     \end{subfigure}
    \begin{subfigure}[b]{0.16\textwidth}
         \centering
         \includegraphics[width=0.8\textwidth, scale=0.5]{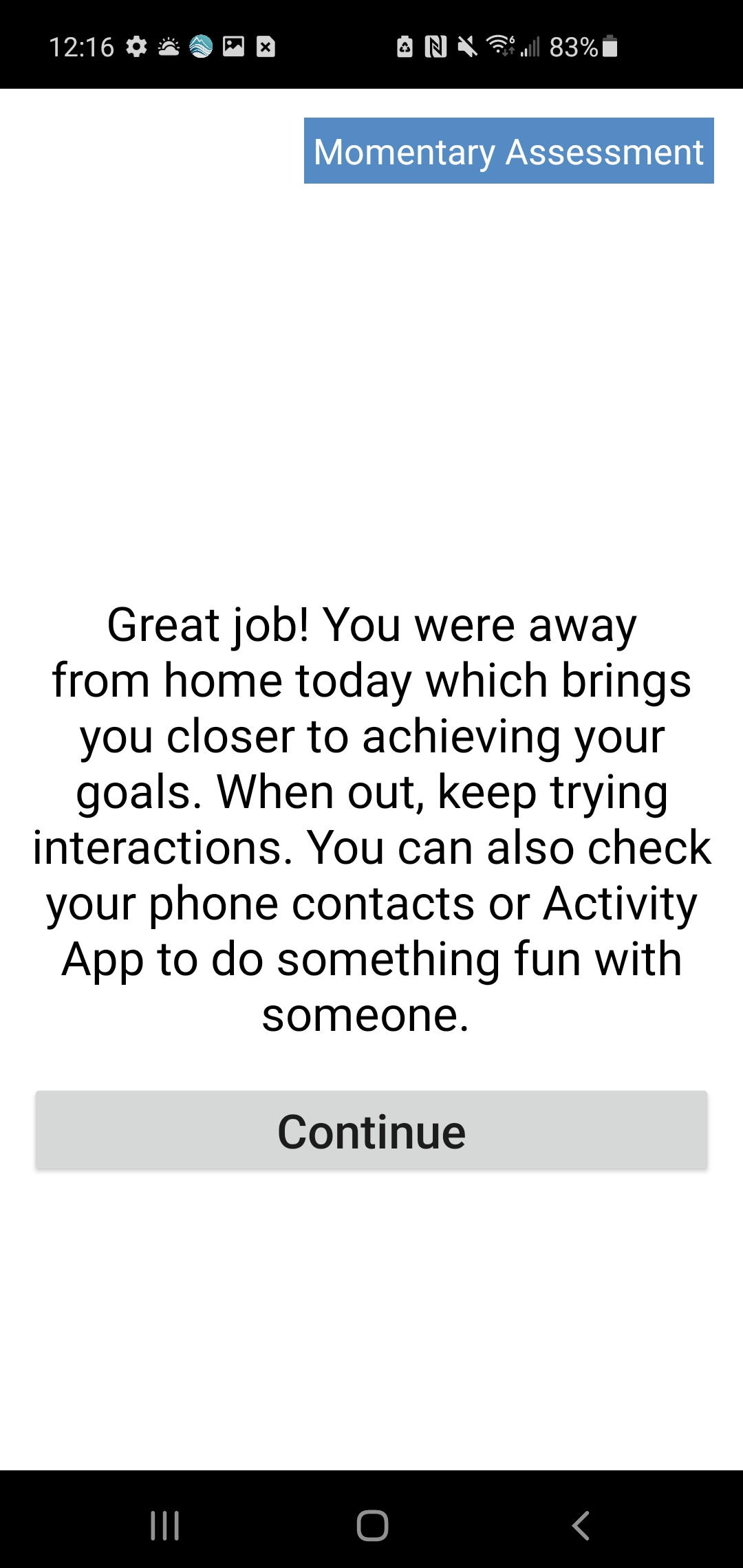}
         \caption{}
         \label{fig:awayalone6}
     \end{subfigure}
      \caption{An example of a context-aware EMA from mSITE. (a) The conversation-location sensor detects that the user is away and alone in the morning. (b)-(d) EMA questions appraising the user's current defeatist or threat beliefs, (e) Challenging the user's defeatist belief through personalization and their conversations with the therapist, (f) A prompt helping the user savor positive actions.}
        \label{fig:msitecontextema}
\end{figure*}
The mSITE app is installed on the participants' phone or an Android phone provided during in-person and coaching treatment phases. Our application consists of five elements:

\subsubsection{Goals} mSITE provides on-demand resources that are populated during individual sessions. Participants set a long-term goal and enter short-term goals and steps to achieve the long-term goal (see Fig. \ref{fig:goalsteps}). They can add their own steps on demand in and out of session. 

\subsubsection{Pleasurable Activities} Participants create a list of enjoyable social activities and rate their anticipated pleasure in doing them (see Fig. \ref{fig:activitypre}). After doing them, they again rate their pleasure and are prompted to do something to savor pleasure (e.g., write about it or use the phone to take a photo or video), which are used as reminders that activities can be more enjoyable than they think (see Fig. \ref{fig:activitypost}).

\subsubsection{Conversation-Location Sensing} As mentioned earlier, our app identifies whether participants are home/away and around conversations to fire contextual EMAs, an example is shown in Fig. \ref{fig:msitecontextema}. To identify the participant's home, we first run DBScan clustering to group significant locations of a participant and then label the location where they spend most of their time from 2 am - 4 am as their home location. We use a geofence tolerance of +/-100 meters to mark whether they are home or away. In addition, the machine learning based conversation detector on the mSITE app identifies whether a participant has been around social interactions. Our application identifies speech and conversation by utilizing pre-trained classifiers that analyze accelerometer and microphone-derived audio data, respectively. To safeguard users' privacy, the application does not retain raw audio data; instead, it stores audio information when speech or conversation is detected. This detection is based on a two-stage classifier: initially, a voice activity detector ascertains the presence of speech, followed by a conversation classifier that determines whether a conversation has commenced and records its duration. Although we cannot be certain if the subject is actively participating in the conversation, the inferred presence of conversation indicates that the subject is in proximity to it. We employ this information as an indicator of social engagement or isolation. The detector analyzes ambient audio every 10 minutes for a duration of one minute to detect the presence of nearby conversations. We tested and refined it to ensure it can distinguish conversations from other sounds such as television and background noise. Our team has previously deployed this conversation detector in numerous studies~\cite{10.1145/2632048.2632054, 10.1145/3313831.3376855}. 

The context-aware scripts using the conversation-location sensor are triggered in afternoon and evening. For example, when the participant is home and alone, the app challenges their negative beliefs about social situations (i.e., anxiety and social threat appraisal) and reminds them of positive experiences they have had in the past. Similarly, when the individual has recently had a social interaction, mSITE helps them savor the positive aspects of that interaction and challenges any negative thoughts they have about future interactions (i.e., defeatist appraisal). This is based on findings from our prior research~\cite{Depp2019, Parrish2020, Depp2016, Granholm2013}; we found that, when home alone, individuals with schizophrenia report threat beliefs and anxiety about going out which contribute to social avoidance, so we challenge threat beliefs. When out around others, they report defeatist beliefs about being judged negatively, so we challenge those beliefs in that context. The app automatically detects whether they are home/away or have been around conversations to fire the personalized interventions. However, we first prompt the participants to confirm that we have indeed correctly detected their social context (e.g., ``I think you have been home and around other people in the afternoon; am I right?"). If they answer with ``no", the app shows ``Sorry that we got it wrong! Were you around others this morning/afternoon?". The challenges that follow are then based on their input to this response rather than on the automatically detected context (i.e., home/away and alone/with others).  
\begin{figure*}[ht!]
\centering
\begin{subfigure}{0.5\textwidth}
  \centering
\includegraphics[width=1\linewidth]{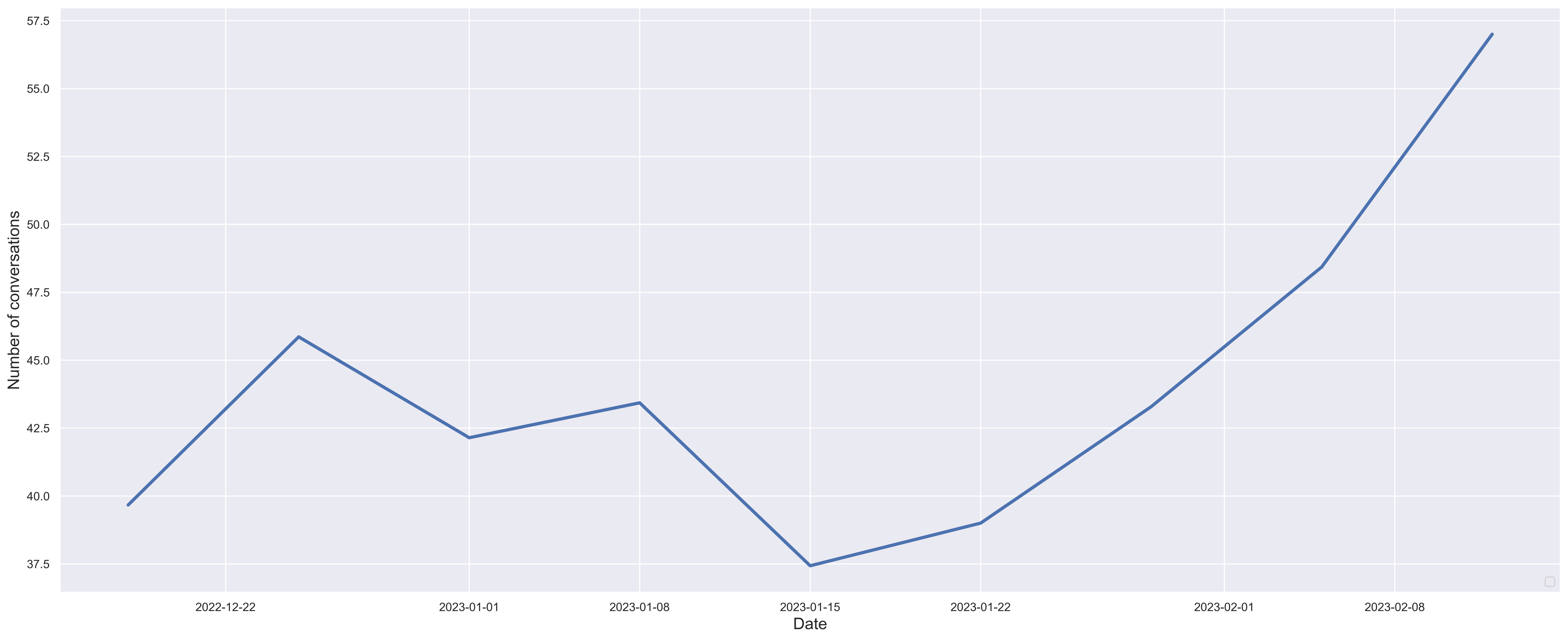}
  \label{fig:stress}
\end{subfigure}%
\hfill
\begin{subfigure}{0.5\textwidth}
  \centering
  \includegraphics[width=1\linewidth]{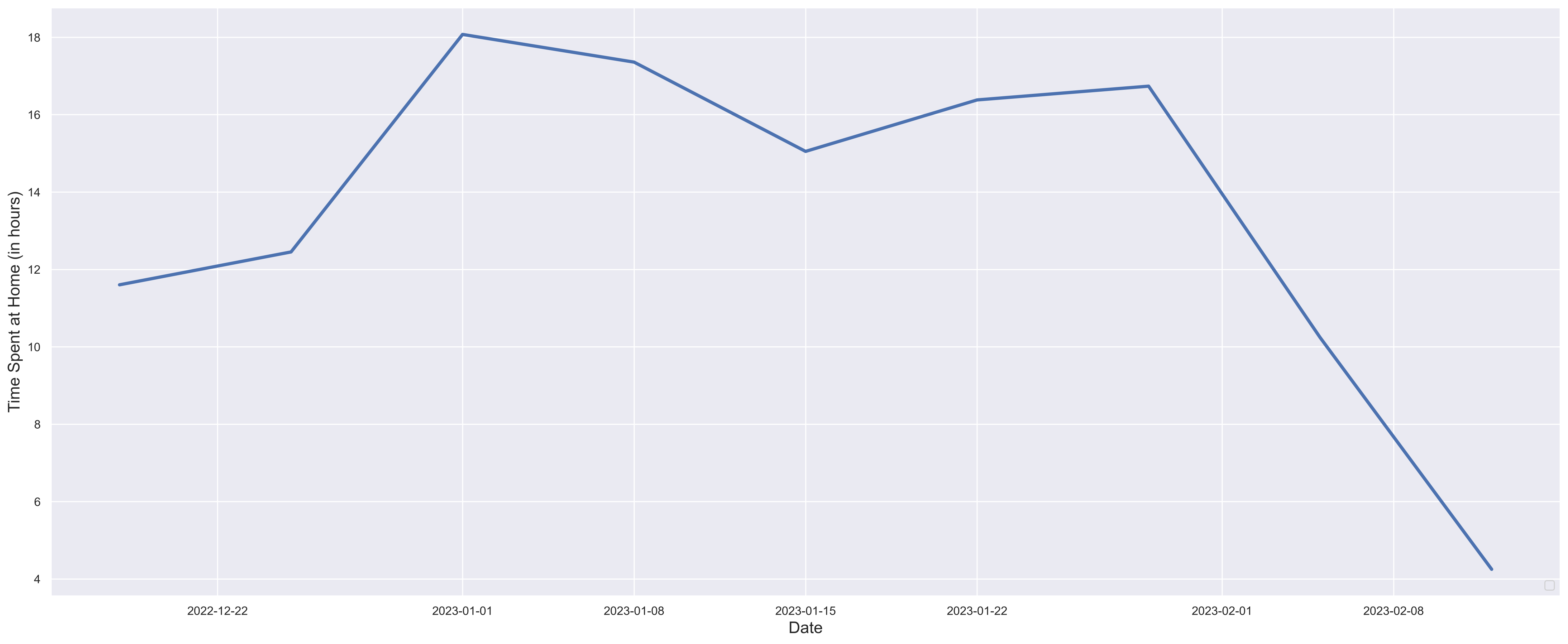}
  \label{fig:anxiety}
\end{subfigure}
\caption{Objectively sensed number of conversations~(left) and time spent at home~(right) of Participant A. We can see that as the study progressed, their number of conversations increased while time spent at home decreased. The x-axis is the date and the y-axis is the sensed data.}
\label{fig:conversation_home}
\end{figure*}
\subsubsection{Therapist-in-the-loop \& personalization} Here, the therapist can enter unlimited personalized messages that challenge the individuals' beliefs using our online therapist dashboard. An example prompt challenging a belief is, ``But it was fun when you played cards with Joe at the clubhouse". For each participant, therapist can add personalized evidence that will challenge either the anxiety beliefs or defeatist beliefs, separately. In addition, the database is composed of several generic messages--a list of pre-composed messages that are not unique to a participant's situation (e.g., ``The best way to find out if someone will like you is to talk to them and see."). The mSITE app randomly selects personalized and generic messages 60\% and 40\% of the times, respectively. The purpose of generic messages are to accommodate new participant who join the study and may have only one or two personalized message. In such cases, using the 60-40 threshold avoid repeated personalized messages. The message is displayed when the context-aware EMAs are fired as shown in Fig. \ref{fig:awayalone5}.

\subsubsection{Gamifying social interactions}
mSITE provides users with an ``Awards" screen that provides information about their intervention progress on demand (see Fig. \ref{fig:awards}). After successful completion of goal steps and activities, an ``awards wheel" randomly gives users ``diamonds". Thus, mildly gamifies the social interaction progress through non-monetary ``achievements".

\section{Application Acceptability}
Our study participants have shown sustained adherence to our mobile app, which is a promising indication of the app's robustness. On average, participants answered 77\% of the EMAs presented to them on their 8 weeks of in-person use. It is worth noting that participants have the option to respond to up to three EMAs daily, including an action plan EMA in the morning and a context-aware EMA in the afternoon and evening. Of the EMAs that participants have answered so far, 46\% are context-aware EMAs triggered by our app's detection of whether the participant is home or away and around conversations. The remaining 54\% of the EMAs are daily action plan EMAs that encourage participants to set goals for their day, such as engaging in a fun activity outside their home, interacting with someone or achieving a custom goal that they can set for themselves. 

Our application demonstrated an initial contextual sensing accuracy of 75\%. To confirm the contextual Ecological Momentary Assessment (EMA) prompt, we have a validation mechanism that solicits participant verification before providing a response. For example, if the application detects that the participant is at home and engaged in conversation, the contextual EMA's initial prompt states, ``I believe you have been at home and in the presence of others this morning; is this correct?" The accuracy of this detection was initially confirmed by participants 75\% of the time. There is a delay between data collection from the application and its integration into the system. For instance, our first contextual EMA triggers at 12 PM, using data collected between 6 AM and 12 PM to determine whether participants have been at home, away, or engaged in conversations. We observed that by adjusting the frequency of data uploads to our server and altering server-side processing intervals, we could enhance accuracy. Since most of our participants live in care homes, we also had to calibrate the voice activity detection model on a noisy environment. In addition, we also played with the geo-fence tolerance to further improve the accuracy. The current version of our application with better calibration has led to an accuracy of over 85\% in internal testing. We are dedicated to continuously refining our settings to further improve the detection accuracy of our system.

In addition to the high adherence rates to EMAs, our mobile sensing data collection has also been successful. Each participant has at least 18 hours of location and audio data available, including information on the presence and amplitude of conversations on 85\% of their study days. Note that we do not record any raw audio. Overall, the acceptability and feasibility of our mobile app-based intervention have been promising, as evidenced by the high adherence rates to both EMAs and mobile sensing data collection. We are continually evaluating the app's effectiveness in improving mental health outcomes and hope to provide additional insights as the study progresses.

\section{Participant Case Study}
\begin{table*}[t]
\caption{8-week study summary for participants A and B evaluating pleasure and interest during interactions, motivation for future interactions, social defeatist attitudes, and time spent at home. The values presented are mean of surveys.}
\label{tab:summary}
\begin{tabular}{@{}lllllll@{}}
\toprule
Participant & Time Point & \begin{tabular}[c]{@{}l@{}}How much pleasure \\ or enjoyment did \\ you feel in the \\ interactions?\end{tabular} & \begin{tabular}[c]{@{}l@{}}In the past hour, \\ how much interest or \\ motivation did you have \\ for interacting with others? \\ {[}Had 0 interactions{]}\end{tabular} & \begin{tabular}[c]{@{}l@{}}How much interest \\ or motivation do you \\ have for engaging in \\ interactions later today? \\ {[}Had 1+ interactions{]}\end{tabular} & \begin{tabular}[c]{@{}l@{}}Social Defeatist \\Attitudes \end{tabular} & \begin{tabular}[c]{@{}l@{}}In the past hour, \\ about how much time \\ did you spend at home?\end{tabular} \\ \midrule

\multirow{2}{*}{A} & Week 0 &3.84 &2.67 &3.16 &3.88 &45.98\\
&Week 8 &4.08 &2.80 &3.68 &3.96 &42.00\\

\multirow{2}{*}{B} & Week 0 &5.39 &2.32 &3.61 &2.48 &54.69\\
&Week 8 &5.45 &1.52 &3.55 &2.36 &53.27\\
\bottomrule
\end{tabular}
\end{table*}

We have enrolled five participants so far for a duration of a little over eight weeks. While it is premature to have any compelling indication about behavior change in this short period, here we highlight two case studies which provides a representative example of the effectiveness of context-aware mobile interventions for improving social functioning. 

Participant A is a 47-year-old White male with 12 years of education. His long-term goal was to make a close friend. While he reported feeling satisfied with the number of acquaintances and social interactions, it was developing a close friendship that was personally meaningful for him. Throughout the in-person sessions, the participant said he enjoyed using the app and found it to be helpful by prompting him each day to do something to either work towards the goal or a task related to it. In the duration of 8 weeks, he answered all the daily action plan EMAs (i.e., 56). The vast majority of the participant's daily action plan is to ``interact with someone" (i.e., 54). The other two instances of action plans are to perform ``fun activity out of home" and ``goal step". In addition, he answered 47 and 40 morning and evening contextual EMAs respectively. He had not interacted with anyone 36\% of the time the contextual EMA was triggered, but the participant had involved in interaction for 64\% of the remaining cases. The contextual sensor also detected that he spent time away from home on most days (over 90\% of the contextual EMAs detect that he was away from home). Since he was out and about most of the time, the personalized therapist messages received are mostly about challenging defeatist believes (e.g., You are more interesting to talk to than you think. Take a chance and have a conversation with someone new today!) and encouraging social interactions (e.g., Making new friends will provide me with companionship). In fact, of the 144 personalized messages the participant received, 46\% of the messages were about defeatist challenges, 44\% that encourage social interactions and the remaining 10\% include messages that challenge anxiety/threat beliefs and encourage making progress towards activities and goals. He appreciated being able to track goal progress in the app and completed numerous goal steps in support of the long-term goal, such as making a list of locations at which the participant could potentially meet new friends, introducing themselves to a new person at one of the locations, and role-playing how to initiate and maintain conversations. He set 10 long-term goals in total, and inserted 25 activities such as doctor visit, go shopping and talk to strangers, attend graduation party and have dinner with family. He has not yet completed the remote-coaching sessions, but has reported feeling happy about his efforts towards meeting new people and greater satisfaction with existing relationships. 

Participant B is a 59-year-old Asian male with 10 years of education. The participant’s initial long-term goal was to make three new friends, which throughout the study later evolved to finding a volunteer position at the local animal shelter. This participant reported that he was comfortable making conversation, but that he would like to become more open with others and to be more involved in the community. Throughout the in-person sessions, the participant had variable engagement with the app, stating that he would leave the phone at home so that he would not lose it. Consequently, during the first 8 weeks, he answered 89\% of the daily action plan EMAs but only 21\% of the contextual EMAs. With coaching, the participant brought the phone with him when he was outside the home and subsequently increased his app engagement and EMA responses. The participant did not use many of the app’s functions such as tracking goal steps and activities, and sometimes struggled to use the phone due to unfamiliarity with the technology. Regardless, the therapist utilized the app’s personalized message function to challenge perception of threat when going out, as well as challenge defeatist attitudes when engaging with others. The participant was engaged with content throughout sessions and he completed many goal steps in support of his long-term goal including starting conversations with others, going to the animal shelter, and filling out an online application. He also increased interactions outside of the home by initiating regular outings with a friend from his board and care. The participant has completed the remote coaching sessions and has made progress toward his goal of volunteering at the local animal shelter.

Fig.~\ref{fig:conversation_home} displays the weekly averages of Participant A's number of conversations and time spent at home, as sensed by the mSITE app. The graph shows that there has been a consistent increase in the number of conversations following the fourth in-person week, and the time spent at home is on a downward trend. From Table \ref{tab:summary}, we observe that pleasure in interactions improved for participants A and B. While participant A's interest in interacting with others improved, participant B lost motivation during past interactions. Both the participants have a decrease in time spent at home. Social defeatist attitudes, on the other hand, increased for participant A, but decreased for participant B. Although eight weeks may be too short to draw definitive conclusions regarding social interaction improvements, the data suggest that the intervention is helping the participant make progress by increasing social interactions and reducing the time spent alone. We hope both the participants will have more movement and interactions by the end of the trial with the app and remote coaching.

\section{DISCUSSION}
\subsection{Preliminary Insights and Implications}
Our preliminary findings demonstrate the feasibility and acceptability of our blended intervention approach that combines in-person CBT with context-triggered mobile CBT interventions to address social isolation in individuals with SMI. Participants have shown sustained adherence to the mobile app, and the app's robustness is evident through its successful data collection and contextual sensing capabilities. As the study progresses, we will continue to evaluate the app's effectiveness in improving social functioning and mental health outcomes in our target population. The context-aware aspect of our intervention is particularly noteworthy, as it sets our approach apart from existing interventions in the field. Most context-aware interventions in the literature have focused on physical health, with limited exploration of their potential in addressing social isolation or mental health outcomes~\cite{ThomasCraig2020}. However, our focus on social activity and functioning is relatively novel in the mental health domain. In fact, it is the very first study to use socially mediated contextual sensing intervention.

The high adherence rates and the case study provided highlight the importance of personalization and therapist involvement in designing mobile interventions. In our study, the mSITE app allowed therapists to input personalized messages that challenge participants' beliefs and tailored interventions. In addition, our mobile intervention system enables users to set their long-term goals and short-term steps, providing them with the tools to work towards achieving those goals. This approach empowers users to take an active role in their recovery and fosters a sense of agency and control, which is crucial for individuals with SMI. Another important implication of our study is the potential for context-aware mobile interventions to be applied in other areas of mental health research. For example, our approach could be used to study social anxiety or agoraphobia, where individuals may be hesitant to leave their homes or interact with others in social situations. In addition, the approach could be used to study healthy aging, where social isolation and loneliness can have negative effects on physical and mental health.

\subsection{Limitations and Future Work}
Our study on social isolation in individuals with SMI has limitations: it's ongoing, requiring more data for conclusions; conducted over a short period; lacks a control group; and focuses on a specific SMI population. In addition, the personalized intervention using mobile sensing technology raises privacy concerns, although no conversation content is recorded. Future research should address these concerns and ensure that the interventions are implemented in a way that respects patient privacy and security. 

Overall, our study represents an important first step in addressing social isolation in individuals with SMI. However, future research is needed to further investigate the effectiveness of context-aware mobile interventions and to address the limitations of our study.

\section{CONCLUSION}
Our study describes a novel blended intervention approach that combines brief in-person CBT with context-triggered mobile CBT interventions to address social isolation in SMI. Our mobile intervention system provides personalized interventions at the right time in the right context to improve social functioning, addressing a critical need in the mental health field. The intervention was acceptable and feasible in this small sample of individuals with SMI. Future research can expand on our work to develop “Just-in-time” adaptive interventions, paving the way for further advancements in the field of mobile sensing and mental health. 

\section{ACKNOWLEDGMENTS}
This work is supported by National Institute of Mental Health (NIMH), grant number 5R61MH126094.
\bibliographystyle{IEEEtran}
\bibliography{referencesFull}

\end{document}